\documentclass[conference]{IEEEtran}
\IEEEoverridecommandlockouts
\usepackage{cite}
\usepackage{amsmath,amssymb,amsfonts}
\usepackage[ruled,linesnumbered]{algorithm2e}
\usepackage{graphicx}
\usepackage{textcomp}
\usepackage{multirow}
\usepackage{xcolor}
\usepackage{booktabs}
\usepackage{algpseudocode}
\usepackage{balance}
\usepackage{listings}
\usepackage{url}
\definecolor{codegreen}{rgb}{0,0.6,0}
\definecolor{codegray}{rgb}{0.5,0.5,0.5}
\definecolor{codepurple}{rgb}{0.58,0,0.82}
\definecolor{backcolour}{rgb}{0.95,0.95,0.92}

\lstdefinestyle{mystyle}{
    backgroundcolor=\color{backcolour},   
    commentstyle=\color{codegreen},
    keywordstyle=\color{magenta},
    numberstyle=\tiny\color{codegray},
    stringstyle=\color{codepurple},
    basicstyle=\ttfamily\footnotesize,
    breakatwhitespace=false,         
    breaklines=true,                 
    captionpos=b,                    
    keepspaces=true,                 
    numbers=left,                    
    numbersep=2pt,                  
    showspaces=false,                
    showstringspaces=false,
    showtabs=false,                  
    tabsize=2
}
\makeatletter
\newcommand{\removelatexerror}{\let\@latex@error\@gobble}
\makeatother

\def\BibTeX{{\rm B\kern-.05em{\sc i\kern-.025em b}\kern-.08em
    T\kern-.1667em\lower.7ex\hbox{E}\kern-.125emX}}
\begin{document}

\title{A Large-Scale Empirical Study on Semantic Versioning in Golang Ecosystem\thanks{\IEEEauthorrefmark{1} Corresponding authors}}

\author{
    \IEEEauthorblockN{Wenke Li\IEEEauthorrefmark{2}\IEEEauthorrefmark{3}, Feng Wu\IEEEauthorrefmark{1}\IEEEauthorrefmark{4}, Cai Fu\IEEEauthorrefmark{1}\IEEEauthorrefmark{2}\IEEEauthorrefmark{3}, Fan Zhou\IEEEauthorrefmark{4}}
    \IEEEauthorblockA{\IEEEauthorrefmark{2} School of Cyber Science and Engineering, Huazhong University of Science and Technology, Wuhan, China}
    \IEEEauthorblockA{\IEEEauthorrefmark{3} Hubei Key Laboratory of Distributed System Security, Hubei Engineering Research Center on Big Data Security}
    \IEEEauthorblockA{\IEEEauthorrefmark{4} Platform and Content Group, Tencent Technology (Shenzhen) Co.Ltd, Shenzhen, China}
    \IEEEauthorblockA{winkli1@hust.edu.cn, barryfwu@tencent.com, fucai@hust.edu.cn, fanzhou@tencent.com}
}

\maketitle

\begin{abstract}

Third-party libraries (TPLs) have become an essential component of software, accelerating development and reducing maintenance costs. However, breaking changes often occur during the upgrades of TPLs and prevent client programs from moving forward. Semantic versioning (SemVer) has been applied to standardize the versions of releases according to compatibility, but not all releases follow SemVer compliance. Lots of work focuses on SemVer compliance in ecosystems such as Java and JavaScript beyond Golang (Go for short). Due to the lack of tools to detect breaking changes and dataset for Go, developers of TPLs do not know if breaking changes occur and affect client programs, and developers of client programs may hesitate to upgrade dependencies in terms of breaking changes.

To bridge this gap, we conduct the first large-scale empirical study in the Go ecosystem to study SemVer compliance in terms of breaking changes and their impact. In detail, we propose \textit{GoSVI (\textbf{G}o \textbf{S}emantic \textbf{V}ersioning \textbf{I}nsight)} to detect breaking changes and analyze their impact by resolving identifiers in client programs and comparing their types with breaking changes. Moreover, we collect the first large-scale Go dataset with a dependency graph from GitHub, including 124K TPLs and 532K client programs. Based on the dataset, our results show that 86.3\% of library upgrades follow SemVer compliance and 28.6\% of no-major upgrades introduce breaking changes. Furthermore, the tendency to comply with SemVer has improved over time from 63.7\% in \textit{2018/09} to 92.2\% in \textit{2023/03}. Finally, we find 33.3\% of downstream client programs may be affected by breaking changes. These findings provide developers and users of TPLs with valuable insights to help make decisions related to SemVer.

\end{abstract}


\section{Introduction}

Third-party libraries (TPLs) nowadays play an important role in modern software systems, avoiding reinventing the wheel and accelerating the development \cite{zhan2021atvhunter,zerouali2019formal,zhan2021research}. Most TPLs provide application programming interfaces (APIs) with downstream client programs to share their new features, fix bugs and refactor code \cite{aline2018apidiff,lina2022breaking}.

However, frequent updates of TPLs may introduce breaking changes and result in compilation errors or runtime errors in client programs, which impose extra costs on client programs to adapt to breaking changes \cite{mostafa2017experience,fan2018large}. In order to communicate with client programs regarding API compatibility, Semantic Version (SemVer) \cite{web:semver} is used by TPLs to signal the state of changes to APIs. SemVer constrains the version number to follow the form of \textit{major.minor.patch.pre-release.build}. The updates of \textit{major} and \textit{pre-release/build} may have incompatible changes (breaking changes), while the updates of \textit{minor} and \textit{patch} compatible changes (no breaking changes).

Unfortunately, not all releases of version numbers are compliant with SemVer. Ochoa et cl. \cite{lina2022breaking} found that in the Maven central, 83.4\% of library upgrades do comply with SemVer, 20.1\% of non-major upgrades introduce breaking changes, and only 7.9\% of breaking changes affect client programs. Decan et al. \cite{alexandrel2021what} proposed an evaluation based on the ``wisdom of crowds'' principle to infer SemVer compliance, and their results showed that packages in Cargo, Npm, and Packagist are more compliant with SemVer than packages in Rubygems. Lots of work studies SemVer in Java, Javascript, PHP, Ruby, and Rust ecosystems, but there is a gap that we do not know the situation of SemVer in the Golang (Go for short) ecosystem.

The Go programming language is becoming increasingly popular among developers as the number of Go repositories starred on GitHub has increased from rank 8 in \textit{Q1/2014} to rank 3 in \textit{Q1/2023} \cite{web:githut}, indicating that more Go TPLs are available. However, not all TPLs adhere to SemVer to meet backward compatibility, for example, the users of popular web framework \textit{gin} found that there were breaking changes in a \textit{minor upgrade} \cite{web:ginissue}. Despite these breaking changes were tagged in the release notes, they still led to confusion for developers of downstream client programs \cite{Raula2017Dodevelopers}. As a result, it is significant to investigate SemVer compliance in the Go ecosystem.

To bridge this gap, in this paper, we perform the first large-scale empirical study to explore SemVer compliance in the Go ecosystem, there are two main challenges.

\begin{itemize}
    \item There is no reliable tool to identify breaking changes and analyze their impact on client programs for Go.
    \item There is no large-scale dataset for Go, which can identify TPLs and corresponding client programs.
\end{itemize}

To address the first challenge, we implement a new tool, \textit{GoSVI (\textbf{G}o \textbf{S}emantic \textbf{V}ersioning \textbf{I}nsight)}, to detect module-based breaking changes and compare usages of client programs with breaking changes based on source code. In detail, first, we extract exported objects from all packages of TPLs, such as interface, function, and struct. Second, we report breaking changes according to 39 conditions based on the Go apidiff tool \cite{web:apidiff}. Finally, we resolve identifiers of client programs to extract their types and compare them with breaking changes to calculate the usage rate.
To address the second challenge, first, we crawl all Go repositories on GitHub with stars greater than five and remove invalid, duplicate, and outdated repositories. Second, The dependency graph is extracted among the repositories by parsing the \textit{go.mod} file, which is stored in the graph database neo4j \cite{web:neo4j}. Finally, we identify TPLs and client programs based on their relationships, where entry nodes are treated as client programs and exit nodes are treated as TPLs.

In our work, we set up three questions to explore SemVer compliance in the Go ecosystem. 

\begin{itemize}
    \item RQ1: How are semantic versioning compliance applied in the Go ecosystem in terms of breaking changes?
    \item RQ2: How much adherence to semantic versioning compliance has increased over time?
    \item RQ3: What about the impact of breaking changes on client programs?
\end{itemize}

In our study, we collect 124K TPLs and 532K client programs according to the dependency graph of our dataset. Our results find that (1) 86.3\% of libraries upgrades follow SemVer compliance, 28.6\% of non-major upgrades (minor and patch upgrades) introduce breaking changes, where \textit{remove} in \textit{function} type is the most common breaking change condition at 36.2\%. (2) the tendency to comply with SemVer has improved over time from 63.7\% in \textit{2018/09} to 92.2\% in \textit{2023/03}. (3) only 3.1\% of breaking changes affect client programs, while 33.3\% of client programs may be affected by breaking changes. Furthermore, we give some suggestions for developers of TPLs and client programs regarding finding breaking changes, maintaining the project layout, and when to upgrade dependencies. Our dataset and tool are accessible on the website \footnote{https://github.com/liwenke1/GoSVI}.

The main contributions of our study are as follows.

\begin{itemize}
    \item We conducted the first large-scale empirical study to analyze SemVer compliance in the Go ecosystem and the impact of breaking changes on client programs. 
    \item we implemented a new tool, \textit{GoSVI}, to detect breaking changes and analyze their impact on client programs.
    \item We proposed a large-scale dataset with a dependency graph in the Go ecosystem, in which TPLs and corresponding client programs are identified for further research.
    \item We offer some suggestions for Go developers of TPLs and client programs from our study of compatibility analysis.
\end{itemize}

\section{Background}

\subsection{Semantic Version}

According to Semantic Version \cite{web:semver}, a version number (x.y.z-pre-release+build) consists of three required parts (Major, Minor, Patch version) and two optional parts (Pre-release and Build Metadata label). Each version change should respect the following rules:

\begin{itemize}
    \item Major version (x) increases when you make incompatible API changes, for example, the version number changes from 1.0.0 to 2.0.0.
    \item Minor version (y) increases when you add functionality in a backwards compatible manner, for example, the version number changes from 1.0.0 to 1.1.0.
    \item Patch version (z) increases when you make backwards compatible bug fixes, for example, the version number changes from 1.0.0 to 1.0.1.
    \item Pre-release label (pre-release) indicates that the version is a pre-release and may not guarantee compatibility requirements compared to the associated normal version, for example, version 1.0.1-alpha may not be compatible with version 1.0.1.
    \item Build metadata label (build) indicates that the version only has different build metadata compared to the associated normal version, for example, version 1.0.1+001 may change build metadata compared to version 1.0.1.
\end{itemize}

Moreover, If the major version is 0, we treat it as the initial development stage which can introduce breaking changes.

\subsection{Go Package Management and Module System}

Package managers are used as tools to create the environment to download, update, and remove project dependencies, allowing developers to build projects more conveniently, such as Maven, Gradle in Java, Npm in JavaScript, and Pip in Python, the Go team also introduced Go modules to manage dependencies \cite{web:go1.11}.

In the module system, a module consists of multiple packages and is used as a unit for managing dependencies \cite{web:gomod-ref}. Each Go module has a \textit{go.mod} file and a \textit{go.sum} file.

\begin{itemize}
    \item The \textit{go.mod} file describes the module’s properties, including module path, its dependencies on other modules, and the Go version.
    \item The \textit{go.sum} file lists down the checksum of direct and indirect dependency required along with the version, which is used to confirm that none of them has been modified.
\end{itemize}

In the package management system, the main process consists of a decentralized package distribution system, a package search engine, a version naming convention, and Go command tools \cite{web:gotools}.

\begin{figure*}[!t]
\centering
\includegraphics[width=\textwidth]{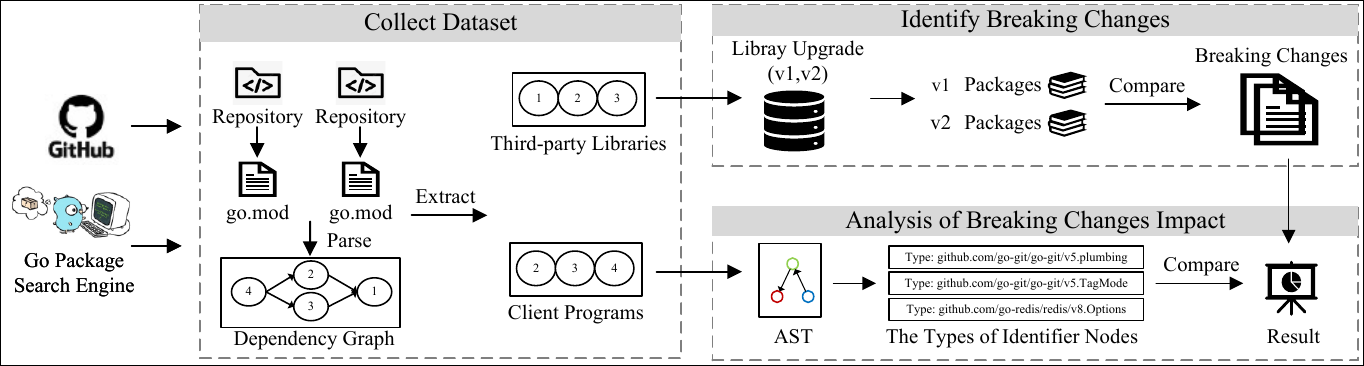}
\caption{Overview of the \textit{GoSVI} tool}
\label{fig:overview}
\vspace{-1em}
\end{figure*}

\begin{itemize}
    \item The decentralised package distribution system allows developers to publish modules in their repositories and download other modules.
    \item Package search engine \cite{web:searchengine} provides developers with the ability to locate and discover useful modules and packages.
    \item Version naming convention helps developers to select the right version in terms of compatibility, it serves the same purpose as SemVer.
    \item Go command tools support an interface to manage dependencies, making it easy for developers to add, update, and remove dependencies.
\end{itemize}

In this paper, we use modules as the unit for detecting breaking changes because modules, which as the unit of the package distribution system, can be imported by client programs. Moreover, we extract the dependency graph by parsing the \textit{go.mod} file of each Go module.

\section{Study Design}

In this section, we describe how we collect the dataset and the implementation of \textit{GoSVI} to detect breaking changes between two valid versions and analyze the impact on client programs, as shown in Figure \ref{fig:overview}.

\subsection{Data Collection} \label{Data Collection}

The purpose of data collection is to get a large number of go modules and the dependency graph among them, which can be used to identify TPLs and client programs. In the package search engine \cite{web:searchengine}, we can locate useful modules with their dependency graph, but there are two limitations:(1) it is difficult to get all modules in the engine; (2) there are many duplicate and deprecated modules in their dependency graph. Therefore, to ensure the confidence and quality of the dataset, we collect the Go repositories from GitHub and extract the dependency graph among them.

\textbf{Crawl Dataset}. We use official Search API \cite{web:githubAPI} published by GitHub to crawl Go repositories. Regarding the limitations on the length of query response, we design two query strategies based on \textit{Star} to collect all the repositories. For repositories with more than 100 stars, keyword \textit{Star} is used to order the query response during pagination. Otherwise, repositories whose \textit{Star} are the same and smaller than 100, will trigger the query response limitations (1,000), resulting in missing some search results, we add another keyword \textit{CreateTime} to order the query response during pagination to solve the problem. Finally, we only reserve repositories with stars greater than five.

\textbf{Clean Dataset}. The repositories, which have many defects, will affect the detection results, so we summarized three situations in which the repositories should be removed from our dataset.

\begin{itemize}
    \item \textbf{Invalid repositories}. Repositories that are empty or have build errors or their version number is smaller than two, can not support detecting breaking changes. We check the extension of every file in each repository, if all files do not equal \textit{"*.go"}, it is considered as an empty repository; we use command \textit{"go build"} in the default branch of the repository to check whether it has build error; we use \textit{"git tag -l"} to calculate the number of versions and remove the repositories whose versions number is smaller than two.
    \item \textbf{Duplicate or similar repositories}. Duplicate repositories are treated as copies of other repositories, as a result, we remove those fork or mirror repositories by searching through GitHub API. Similar repositories, defined as having cloned, migrated, or deprecated repositories, will have the same module path and version number list, so we keep the latest repositories based on the update time.
    \item \textbf{Repositories with Go version before 1.11}. We remove these repositories because Go 1.11 adds preliminary support for a new concept called "modules", an alternative to \textit{GOPATH} with integrated support for versioning and package distribution, and support locating and loading packages of Go source code, what's more, the dependence is defined and managed with \textit{go.mod} file. We remove the repositories that do not contain \textit{go.mod} and have errors during loading and resolving packages.
\end{itemize}

\textbf{Extract Dependency Graph}. The purpose of extracting the dependency graph is to identify TPLs and their corresponding client programs in the dataset. To this end, we parse the \textit{go.mod} file of each repository, analyze the \textit{module} field and the \textit{require} field, and obtain the module path and dependency information on other modules, including the module paths, version numbers, and relationship. We filter out all modules that are not directly imported since no package from these modules is called directly. Note that multiple modules may exist in a repository, so each module will be parsed separately and stored separately. Finally, the dependency graph is stored in the graph database \textit{neo4j} \cite{web:neo4j}, where different versions of each module are treated as nodes and the dependency relationships are treated as edges. In the dependency graph, nodes will be identified as TPLs only if they are dependent on other modules, and their downstream nodes are treated as client programs.

\subsection{Breaking Change Detection} \label{Breaking Change Detection}

\begin{figure}
\centering
\begin{lstlisting}[]
Module: github.com/pinpoint-apm/pinpoint-go-agent
Library Upgrade: v1.1.3 -> v1.2.0, Minor Upgrade
Package: github.com/pinpoint-apm/pinpoint-go-agent/protobuf
Change Node: NewAgentClient
Change Category: Function
Change Condition: Param Change
Change Message: func(google.golang.org/grpc.ClientConnInterface) AgentClient -> func(*google.golang.org/grpc.ClientConn) AgentClient
\end{lstlisting}
\vspace{-1em}
\caption{The Output of Identifying Breaking Changes}
\label{fig:output}
\vspace{-1em}
\end{figure}

In our work, we utilize and enhance the Go apidiff tool \cite{web:apidiff}, proposed by Go maintainers, to detect breaking changes between versions of TPLs. In this section, we introduce the methodology of detecting breaking changes and the catalogue of breaking changes.

The Go apidiff tool is designed to display the differences in exported objects between any two packages, including compatibility and incompatibility changes (breaking changes). We use it to detect breaking changes, however, there are two limitations:(1) the brief report messages make it difficult to classify the results for further research; (2) it cannot detect breaking changes between modules. To address the first limitation, we optimized the tool to report the location and type of breaking changes, as shown in Figure \ref{fig:output}, including module, package, type, condition, and detailed message of change nodes. To address the second limitation, we parse all packages of the modules, detect breaking changes between packages with the same path, and finally merge all results. Moreover, the detection process is built completely automatically on repositories hosted on git workflows.

Table \ref{tab:BCs} presents the catalogue of breaking changes. There are 13 categories of breaking changes, consisting of \textit{Package} change, 11 data types changes in Go syntax, and \textit{Data Type Change}, which is defined as the change between the above 11 data types. 
It should be noted that for \textit{basic} date type, we set \textit{basic (Const)} because of its extra breaking change condition of \textit{value change}. A data type may be declared in multiple scenarios, for example, \textit{Map} can be declared by using the keywords var or type. Furthermore, we summarize 39 conditions of breaking changes based on the change, removal, and addition of its components. Note that \textit{Add Unexported Method} is adding a method to an interface without an existing unexported method, which will make a type that implements the old interface fail to implement the new one. 

\subsection{Analysis of Breaking Changes Impact} \label{Analysis of Breaking Changes Impact}

To measure the impact of breaking changes on the client programs, we parse the ASTs of packages in the client programs based on static analysis and resolve identifiers to obtain their types. By comparing the breaking change nodes with the identifier nodes of client programs, we are able to get the results of the impact analysis.

Algorithm \ref{alg:Impact} gives the process of analyzing the impact on client programs. First, for each TPL, we extract the type list $t(n)$ and package path list $p(n)$ of the breaking change nodes $n$ (line 2-4). Second, we extract the package path list $ p(c)$ of client $c$, and by comparing $p(b)$ and $p(c)$, we filter out client programs that do not import packages of breaking change nodes (lines 5-9). Otherwise, we resolve each identifier node $n$ to obtain its type $t(n)$ by parsing the ASTs of the client program, when the type of the identifier node $t(n)$ is in the type list $t(n)$ of breaking change nodes, it means that the client program will be affected by this breaking change (line 10-15).

\begin{table}[!t]
\caption{catalogue Of Breaking Changes In The Go Ecosystem}
\label{tab:BCs}
\centering
\resizebox{0.95\linewidth}{!}{
\begin{tabular}{ll}
\bottomrule
\textbf{Category} & \textbf{Condition}                                                                                                                                          \\ \hline
Package           & Remove                                                                                                                                                      \\
Basic (Const)     & Type Change, Value Change, Remove                                                                                                                           \\
Basic  & Type Change, Remove                                                                                                                                         \\
Array             & Element Change, Length Change, Remove                                                                                                                       \\
Slice             & Element Change, Remove                                                                                                                                      \\
Map               & Key Change, Value Change, Remove                                                                                                                            \\ \hline
Struct            & \begin{tabular}[c]{@{}l@{}}Field Number Change, Field Anonymous Change,\\ Field Type Change, Field Name Change,\\ Field Tag Change, Comparability Change, Remove\end{tabular} \\ \hline
Interface         & \begin{tabular}[c]{@{}l@{}}Method Number Change, Method ID Change,\\ Add Unexported Method, Add Interface Method,\\ Remove\end{tabular}                     \\ \hline
Pointer           & Base Change, Remove                                                                                                                                         \\
Channel              & Element Change, Direction Change                                                                                                                            \\ \hline
Function          & \begin{tabular}[c]{@{}l@{}}Param Change, Return Change,\\ Variadic Change, Remove\end{tabular}                                                              \\ \hline
Named             & Element Change, Remove                                                                                                                                      \\
TypeParam         & Type Change, Remove                                                                                                                                         \\
Data Type Change        & Data Type Change                                                                                                                                            \\ \bottomrule
\end{tabular}
}
\vspace{-1em}
\end{table}

\begin{figure}[!t]
\renewcommand{\algorithmicrequire}{\textbf{Input:}}
\renewcommand{\algorithmicensure}{\textbf{Output:}}
\removelatexerror
\begin{algorithm}[H]
    \label{alg:Impact}
	\caption{Algorithm of Impact Analysis}
	\LinesNumbered
	\KwIn{$L$: third-party libraries. $C(i)$: clients for third-party library $i$. $B(i)$: nodes of breaking changes in third-party library $i$.}
	\KwOut{$R$: affected nodes in the clients}
        $R \gets \phi$\;
        \ForEach{{\rm third-party library} $l$ {\rm in} $L$}{
            $t(b) \gets$ types of $B(l)$\;
            $p(b) \gets$ package paths of $B(l)$\;
            \ForEach{{\rm client} $c$ {\rm in} $C(l)$}{
                $p(c) \gets $ import package paths of $c$\;
                \If{$p(b) \cap p(c) = \phi$}{
                    continue\;
                }
                \ForEach{{\rm Identifier Node} $n$ {\rm in AST of} $c$}{
                    $t(n) \gets Resolve(n)$\;
                    \If{$t(b) \cap t(n) \ne \phi$}{
                        $R \gets \left \langle c, n \right \rangle$\;
                    }
                }
            }
        }
        \Return return $R$
\end{algorithm}
\vspace{-1em}
\end{figure}

\section{Study Results}

In this section, we present the scope of our dataset and answer each question through experiments and analysis. All experiments were performed under Go version 1.19.6.

\subsection{Dataset Scope}

Table \ref{tab:scope} shows the results of crawling and cleaning the dataset in different dimensions. There are 102,420 repositories with stars greater than five, we filter out invalid repositories according to the rule proposed in Section \ref{Data Collection} and obtain 29,095 valid repositories. For the remaining repositories with 559,693 versions, we filter out all versions that do not meet the SemVer constraint because we cannot analyze their compatibility. Furthermore, we classify the valid versions into five categories based on the type of library upgrade, including \textit{major upgrade} (6,691), \textit{minor upgrade} (45,634), \textit{patch upgrade} (118,912), \textit{development} (200,859), and \textit{pre-release/build} (75,084). \textit{development}, with a major version number of 0, is considered to be the initial development stage and can introduce breaking changes like \textit{major upgrade} and \textit{pre-release/build}. \textit{minor upgrade} and \textit{patch upgrade} cannot introduce breaking changes to ensure compatibility with client programs. We find that \textit{development} is the most frequent of the five library upgrades, meaning that there are more Go repositories in the unstable development stage. Finally, we identified the TPLs and corresponding client programs based on the dependency graph, there are 5,604 TPLs with 124,532 versions and 23,929 client programs with 532,832 versions respectively. Note that the versions of the TPLs must meet the SemVer constraint, while the versions of the client programs do not.

\begin{table}[!t]
\caption{Scope Of Dataset}
\label{tab:scope}
\centering
\resizebox{0.95\linewidth}{!}{
\begin{tabular}{llllr}
\bottomrule
\textbf{Dimension}          & \textbf{Validty}       & \multicolumn{2}{l}{\textbf{Category}}          & \textbf{Number} \\ \hline
\multirow{3}{*}{Repository} & Total                  & \multicolumn{2}{l}{}                  & 102,420         \\
                            & Invalid                & \multicolumn{2}{l}{}                  & 73,325          \\
                            & Valid                  & \multicolumn{2}{l}{}                  & 29,095          \\ \hline
\multirow{8}{*}{Version}    & Total                  & \multicolumn{2}{l}{}                  & 559,693         \\
                            & Invalid                & \multicolumn{2}{l}{}                  & 112,513         \\ \cline{2-5} 
                            & \multirow{6}{*}{Valid} & \multicolumn{2}{l}{Total}             & 447,180         \\
                            &                        & \multicolumn{2}{l}{Major Upgrade}     & 6,691           \\
                            &                        & \multicolumn{2}{l}{Minor Upgrade}     & 45,634          \\
                            &                        & \multicolumn{2}{l}{Patch Upgrade}     & 118,912         \\
                            &                        & \multicolumn{2}{l}{Development}       & 200,859         \\
                            &                        & \multicolumn{2}{l}{Pre-release/Build} & 75,084          \\ \hline
\multirow{4}{*}{Dependency Graph}     & \multirow{4}{*}{Valid} & \multirow{2}{*}{TPL}     & Repository & 5,604           \\
                            &                        &                          & Version    & 124,532          \\ \cline{3-5} 
                            &                        & \multirow{2}{*}{Client}  & Repository & 23,929          \\
                            &                        &                          & Version    & 532,832         \\ \bottomrule
\end{tabular}
}
\vspace{-1em}
\end{table}

\begin{table}[!t]
\caption{Total And Breaking Upgrades In The Dataset}
\label{tab:BCInUpgrade}
\centering
\begin{tabular}{llrrrr}
\bottomrule
\multicolumn{2}{l}{\multirow{2}{*}{\textbf{Level}}} & \multicolumn{2}{c}{\textbf{Total}} & \multicolumn{2}{c}{\textbf{Breaking}} \\ \cline{3-6} 
\multicolumn{2}{l}{}                                & \textbf{Count}    & \textbf{\%}    & \textbf{Count}     & \textbf{\%}    \\ \hline
\multicolumn{2}{l}{Repository}                      & 5,604             & 100            & 1,674              & 29.9             \\ \hline
\multirow{5}{*}{Version}        & Major             & 1,926             & 1.5            & 1,147              & 59.6              \\
                                & Minor             & 16,305            & 13.1           & 6,173              & 37.9             \\
                                & Patch             & 43,222            & 34.7           & 10,836             & 25.1             \\
                                & Development       & 63,079            & 50.7           & 17,904             & 28.4             \\
                                & Non-Major         & 59,527            & 47.8           & 17,009             & 28.6             \\ \cline{2-6}
\multicolumn{1}{c}{}            & Total             & 124,532           & 100            & 36,060             & 29.0              \\ \bottomrule
\end{tabular}
\vspace{-1em}
\end{table}

\begin{figure}[!t]
\centering
\includegraphics[width=0.47\textwidth]{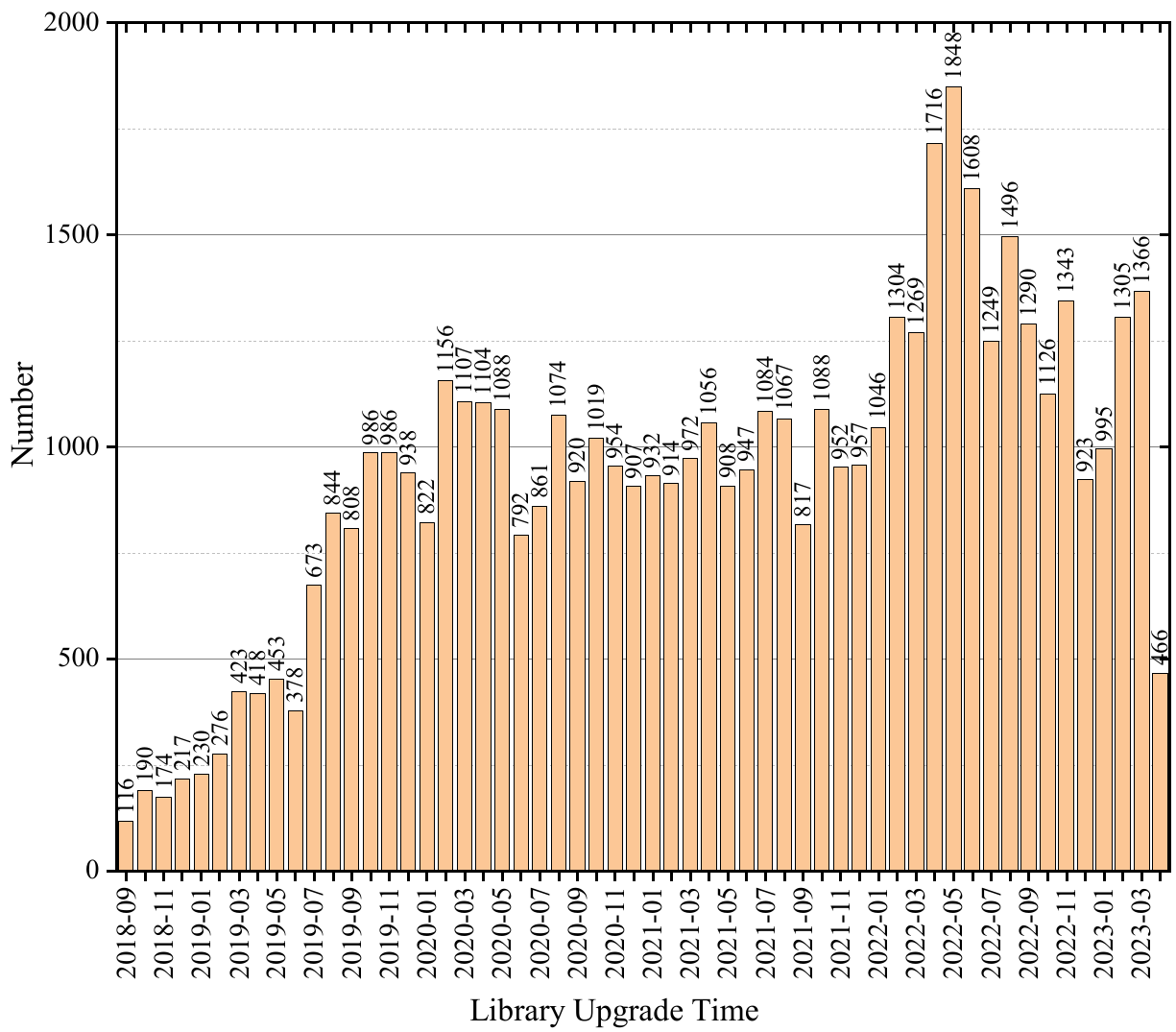}
\caption{Statistics on the number of version upgrades over time}
\label{fig:LibraryVersionCreateTime}
\vspace{-1em}
\end{figure}

We also present the distribution of valid library upgrade numbers ranging from \textit{2018/09} to \textit{2022/04}, as shown in Figure \ref{fig:LibraryVersionCreateTime}. We chose the start date of \textit{2018/09} because the Go package management and module system was launched in Go version 1.11 in \textit{2018/09} \cite{web:go1.11}. In \textit{2023/04}, the number of library upgrade number drops sharply relative to the previous date since we collect repositories with stars greater than five before \textit{2023/04} and recent repositories have less time to get five stars. Overall, the number of library upgrade number is increasing from 116 in \textit{2018/09} to 1,366 in \textit{2023/03}. Therefore, with the rapid increase in library upgrades, it is meaningful to detect whether library upgrades meet SemVer compliance and do not introduce breaking changes.

\begin{center}
\noindent\fbox{
\parbox{0.45\textwidth}{
\textit{Summary:} We identified 124K TPLs and 532K client programs from the collected dataset. Moreover, the number of valid library upgrades is rapidly increasing over time and it is significant to explore compliance with SemVer.
}
}
\end{center}

\subsection{RQ1: How are semantic versioning compliance applied in the Go ecosystem in terms of breaking changes}

\subsubsection{Method}

With RQ1, we analyze the extent of SemVer principles applied in the Go ecosystem from three perspectives: breaking repositories, breaking library upgrades, and the types of breaking changes. First, we identify how many repositories have introduced breaking changes. Second, we analyze how many breaking changes are usually introduced in each library upgrade to know the distribution of breaking changes. Finally, we study the frequency of each type of breaking change to find out the most common breaking changes in our dataset.

\begin{table*}[!t]
\caption{The Distribution Of Breaking Changes And Their Impact}
\label{tab:distribution}
\centering
\begin{tabular}{l|llrrrrrrr}
\bottomrule
\multirow{2}{*}{\textbf{Index}} & \multirow{2}{*}{\textbf{Category}} & \multirow{2}{*}{\textbf{Condition}} & \multicolumn{2}{c}{\textbf{Breaking Change}} & \multicolumn{3}{c}{\textbf{\begin{tabular}[c]{@{}c@{}}Usage in \\ Client Program\end{tabular}}} & \multicolumn{2}{c}{\textbf{\begin{tabular}[c]{@{}c@{}}Breaking\\ Library Upgrade\end{tabular}}} \\ \cline{4-10} 
                                &                                    &                                     & \textbf{Number (B)}     & \textbf{\%}   & \textbf{Number (U)} & \textbf{\%}       & \textbf{\% (U/B)}                        & \textbf{Number}                                & \textbf{\%}                               \\ \hline
1                               & Package                            & Remove                              & 4,132               & 1.1              & 165                    & 1.5                & 4.0                      & 2,091                                          & 4.1                                       \\ \hline
2                               & \multirow{3}{*}{Basic (Const)}     & Type Change                          & 1,332               & 0.4              & 22                     & 0.2               & 1.7                       & 53                                             & 0.1                                       \\ \cline{1-1}
3                               &                                    & Value Change                         & 26,955              & 7.4              & 594                    & 5.3               & 2.2                       & 18,788                                         & 37.2                                      \\ \cline{1-1}
4                               &                                    & Remove                              & 22,738              & 6.3              & 529                    & 4.7                & 2.3                      & 265                                            & 0.5                                       \\ \hline
5                               & \multirow{2}{*}{Basic}  & Type Change                          & 1,608               & 0.4              & 1                      & 0.0               & 0.0                       & 15                                             & 0.0                                       \\ \cline{1-1}
6                               &                                    & Remove                              & 13,223              & 3.6              & 3                      & 0.0                & 0.0                      & 20                                             & 0.0                                       \\ \hline
7                               & \multirow{3}{*}{Array}             & Element Change                       & 12                  & 0.0              & 0                      & 0                 & 0                       & 0                                              & 0                                         \\ \cline{1-1}
8                               &                                    & Length Change                        & 60                  & 0.0              & 0                      & 0                 & 0                       & 0                                              & 0                                         \\ \cline{1-1}
9                               &                                    & Remove                              & 73                  & 0.0              & 0                      & 0                  & 0                      & 0                                              & 0                                         \\ \hline
10                              & \multirow{2}{*}{Slice}             & Element Change                       & 928                 & 0.3              & 3                      & 0.0               & 0.3                       & 10                                             & 0.0                                       \\ \cline{1-1}
11                              &                                    & Remove                              & 3,709               & 1.0                & 0                      & 0                & 0                        & 0                                              & 0                                         \\ \hline
12                              & \multirow{3}{*}{Map}               & Key Change                           & 63                  & 0.0              & 0                      & 0                 & 0                       & 0                                              & 0                                         \\ \cline{1-1}
13                              &                                    & Value Change                         & 247                 & 0.1              & 0                      & 0                 & 0                       & 0                                              & 0                                         \\ \cline{1-1}
14                              &                                    & Remove                              & 1,350               & 0.4              & 6                      & 0.1                & 0.4                      & 8                                              & 0.0                                       \\ \hline
15                              & \multirow{7}{*}{Struct}            & Field Number Change                   & 195                 & 0.1              & 3                      & 0.0              & 1.5                        & 3                                              & 0.0                                       \\ \cline{1-1}
16                              &                                    & Field Anonymous Change                & 0                   & 0                & 0                      & 0                & 0                        & 0                                              & 0                                         \\ \cline{1-1}
17                              &                                    & Field Type Change                     & 140                 & 0.0              & 1                      & 0.0              & 0.7                        & 1                                              & 0.0                                       \\ \cline{1-1}
18                              &                                    & Field Name Change                     & 6                   & 0.0              & 0                      & 0                & 0                        & 0                                              & 0                                         \\ \cline{1-1}
19                              &                                    & Field Tag Change                      & 48                  & 0.0              & 0                      & 0                & 0                        & 0                                              & 0                                         \\ \cline{1-1}
20                              &                                    & Comparability Change                 & 4,567               & 1.3              & 367                    & 3.2               & 8.0                       & 8,325                                          & 16.5                                      \\ \cline{1-1}
21                              &                                    & Remove                              & 970                 & 0.3              & 0                      & 0                  & 0                      & 0                                              & 0                                         \\ \hline
22                              & \multirow{5}{*}{Interface}         & MethodNumber Change                  & 4                   & 0.0              & 0                      & 0                 & 0                       & 0                                              & 0                                         \\ \cline{1-1}
23                              &                                    & Method ID Change                      & 2                   & 0.0              & 0                      & 0                & 0                        & 0                                              & 0                                         \\ \cline{1-1}
24                              &                                    & Add Unexported Method                 & 101                 & 0.0              & 0                      & 0                & 0                        & 0                                              & 0                                         \\ \cline{1-1}
25                              &                                    & Add Interface Method                  & 30,563              & 8.4              & 2,660                  & 23.5             & 8.7                        & 22,899                                         & 45.4                                      \\ \cline{1-1}
26                              &                                    & Remove                              & 102                 & 0.0              & 0                      & 0                  & 0                      & 0                                              & 0                                         \\ \hline
27                              & \multirow{2}{*}{Pointer}           & Base Change                          & 969                 & 0.3              & 2                      & 0.0               & 0.2                       & 170                                            & 0.3                                       \\ \cline{1-1}
28                              &                                    & Remove                              & 4,968               & 1.4              & 1                      & 0.0                & 0                      & 6                                              & 0.0                                       \\ \hline 
29                              & \multirow{3}{*}{Channel}              & Element Change                       & 15                  & 0.0              & 0                      & 0              & 0.0                          & 0                                              & 0                                         \\ \cline{1-1}
30                              &                                    & Direction Change                     & 0                   & 0                & 0                      & 0                 & 0                       & 0                                              & 0                                         \\ \cline{1-1}
31                              &                                    & Remove                              & 71                  & 0.0              & 0                      & 0                  & 0                      & 0                                              & 0                                         \\ \hline
32                              & \multirow{4}{*}{Function}          & Param Change                         & 37,864              & 10.4              & 786                    & 7.0              & 2.1                        & 8,141                                          & 16.1                                      \\ \cline{1-1}
33                              &                                    & Return Change                        & 17,351              & 4.8              & 418                    & 3.7               & 2.4                       & 1,862                                          & 3.7                                       \\ \cline{1-1}
34                              &                                    & Variadic Change                      & 3,917               & 1.1              & 193                    & 1.7               & 4.9                       & 5,054                                          & 10.0                                      \\ \cline{1-1}
35                              &                                    & Remove                              & 131,565             & 36.2             & 829                    & 7.3                & 0.6                      & 2,329                                          & 4.6                                       \\ \hline
36                              & \multirow{2}{*}{Named}             & Element Change                       & 10,039              & 2.8              & 3,783                  & 33.5              & 37.7                       & 1,254                                          & 2.5                                       \\ \cline{1-1}
37                              &                                    & Remove                              & 35,701              & 9.8              & 853                    & 7.5                & 2.4                      & 1,469                                          & 2.9                                       \\ \hline
38                              & \multirow{2}{*}{TypeParam}         & Type Change                          & 21                  & 0.0              & 0                      & 0                 & 0                       & 0                                              & 0                                         \\ \cline{1-1}
39                              &                                    & Remove                              & 0                   & 0                & 0                      & 0                  & 0                      & 0                                              & 0                                         \\ \hline
40                              & Category Change                     & Data Type Change                      & 7,819               & 2.2              & 85                     & 0.8             & 1.1                         & 1,012                                          & 2.0                                       \\ \hline
41                              & Total                              &                                     & 363,428             & 100              &11,304                 & 100                  & 3.1                   & 50,485                                         & 100                                     \\ \bottomrule
\end{tabular}
\vspace{-1em}
\end{table*}

To do so, we identify breaking changes between two valid versions of TPLs and count the types of breaking changes. The types of library upgrades are distinguished into \textit{major upgrade}, \textit{minor upgrade}, \textit{patch upgrade}, \textit{development}, and \textit{non-major upgrade}, ignoring \textit{pre-release/build} because it is usually complementary to the normal library upgrade. \textit{non-major upgrade} consists of \textit{minor upgrade} and \textit{patch upgrade} and cannot introduce breaking changes based on SemVer compliance.

According to the layout of Go projects \cite{web:layout}, not all packages are imported by client programs. Therefore, We filter out the following folders that are not related to breaking changes.

\begin{itemize}
    \item \textbf{/cmd:} Main applications for this project.
    \item \textbf{/internal:} Private application and library code that do not want others importing in their applications or libraries.
    \item \textbf{/vendor:} Application dependencies.
    \item \textbf{/config:} Configuration file templates.
    \item \textbf{/init:} System initialization and management files.
    \item \textbf{/scripts:} Scripts to perform build, install, etc. operations.
    \item \textbf{/build:} Packaging and continuous integration.
    \item \textbf{/deployment:} Deployment configuration files.
    \item \textbf{/test:} Additional external test applications and data.
\end{itemize}

\subsubsection{Results}

From the perspective of breaking repositories, as shown in Table \ref{tab:BCInUpgrade}, 1,674 repositories (29.9\%) have at least one breaking change, meaning that almost one-third of TPLs violate the SemVer compliance and there is a high risk for client programs to pick dependencies that do not adhere to compatibility constraints.

\begin{figure*}[!t]
\centering
\includegraphics[width=0.95\textwidth]{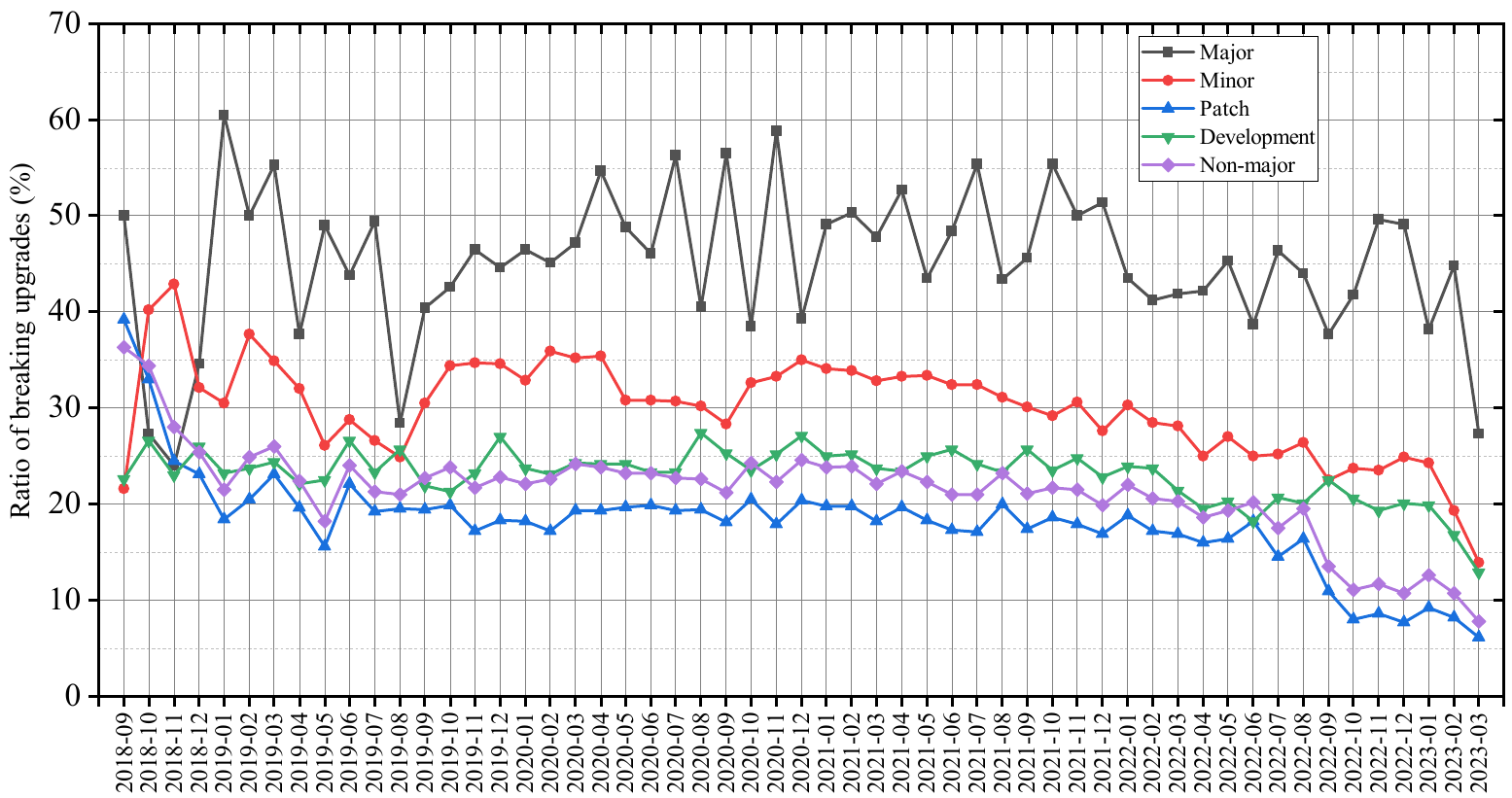}
\caption{SemVer compliance in the Go ecosystem over time}\label{fig:bcovertime}
\vspace{-1em}
\end{figure*}

From the perspective of breaking library upgrades, we found that 50.7\% of library upgrades are in the initial development stage, followed by \textit{patch upgrades} (34.7\%), \textit{minor upgrades} (13.1\%), and \textit{major upgrades} (1.5\%). This distribution indicates that half of the library upgrades are not ready for the public API. For breaking library upgrades, as expected, most \textit{major upgrades} (59.6\%) introduce breaking changes, followed by \textit{minor upgrades} (37.9\%), \textit{development} (28.4\%), \textit{patch upgrades} (25.1\%). There are 28.6\% of  \textit{non-major upgrades} that do not meet the SemVer compliance, as many as the breaking repositories, meaning that one-third of library upgrades introduce breaking changes. It is also important to note that the rate of breaking library upgrades in \textit{development} differs significantly from \textit{major upgrade} but only slightly from \textit{non-major upgrade}, pointing to greater attention to semantic version compliance during the initial development stage.

From the perspective of the types of breaking changes, as shown in Table \ref{tab:distribution}, there are 40 breaking change conditions, divided into 14 categories, including \textit{package}, \textit{basic (const)}, \textit{basic}, \textit{array}, \textit{slice}, \textit{map}, \textit{struct}, \textit{interface}, \textit{pointer}, \textit{channel}, \textit{function}, \textit{named}, \textit{typeParam}, and \textit{category change}. We identified 363,428 breaking changes, and the five most common breaking change types are \textit{remove} in \textit{function} type (36.2\%), \textit{param change} in \textit{function} type (10.4\%), \textit{remove} in \textit{named} type (9.8\%), \textit{add interface method} in \textit{interface} type (8.4\%), and \textit{value change} in \textit{basic (const)} type (7.4\%). Moreover, for breaking change actions such as remove, change, and add, we found that remove is the most common action (60.2\%). 

\begin{center}
\noindent\fbox{
\parbox{0.45\textwidth}{
\textit{Summary:} 86.3\% of library upgrades adhere to SemVer compliance, but 28.6\% \textit{non-major upgrades} introduce breaking changes. It is important to note that even in the initial development stage, developers of TPLs pay more attention to avoid introducing breaking changes relative to \textit{Major Upgrades}. Moreover, remove is the most common breaking change action at 60.2\% while \textit{remove} in \textit{function} is the most commons breaking change condition at 36.2\%.
}
}
\end{center}

\subsection{RQ2: How much adherence to semantic versioning compliance has increased over time?}

\subsubsection{Method}

To answer RQ2, we study the tendency of compliance with SemVer over time in terms of library upgrades. In detail, library upgrades are still divided into five categories, including \textit{major upgrade}, \textit{minor upgrade}, \textit{patch upgrade}, \textit{development}, and \textit{non-major upgrade}. Moreover, We count the rate of breaking library upgrades per month to reflect the tendency of SemVer compliance over time. We do not study the tendency of compliance with SemVer over time in terms of repositories because too few repositories may lead to high data volatility while the number of library upgrades is nearly 22 times the number of repositories.

\subsubsection{Results}

Figure \ref{fig:bcovertime} shows the tendency of the rate of different breaking library upgrades in the latest five years, from \textit{2018/09} to \textit{2023/03}. We focus on the tendency of the rate of breaking \textit{non-major upgrade} because this upgrade cannot introduce breaking changes. Also, We divide the tendency into the following three parts.

\begin{itemize}
    \item \textbf{From \textit{2018/09} to \textit{2019/01}}. Due to the recent release of the Go module system and the low number of library upgrades, the rates of breaking library upgrades change a lot during this period. However, the rate of breaking \textit{non-major upgrades} is decreasing from 36.3\% to 21.5\%.
    \item \textbf{From \textit{2019/01} to \textit{2022/08}}. The rates of breaking library upgrades tend to be stable during this period, where the rate of breaking \textit{non-major upgrades} ranges from 17.5\% to 26.0\%. Regarding \textit{major upgrade}, the rate of it changes a lot because of its small number.
    \item \textbf{From \textit{2022/08} to \textit{2023/03}}. The rates of all breaking library upgrades are decreasing, in which the rate of breaking \textit{non-major upgrade} is decreasing sharply from 19.5\% to 7.8\%.
\end{itemize}

\begin{center}
\noindent\fbox{
\parbox{0.45\textwidth}{
\textit{Summary:} The rate of breaking \textit{major upgrade} fluctuates over time because of its small number, while the rates of other breaking library upgrades are decreasing. Furthermore, more library upgrades meet SemVer compliance from 63.7\% in \textit{2018/09} to 92.2\% in \textit{2023/03}.
}
}
\end{center}

\subsection{RQ3: What about the impact of breaking changes on client programs?}

\subsubsection{Method}

In RQ3, we examine to what extent breaking changes affect client programs in terms of the types of breaking changes and how many client programs are affected by breaking changes. In detail, for a breaking library upgrade from version \textit{v1} to \textit{v2}, we extract the client programs that depend on version \textit{v1}. If the developers of the client programs update the dependent version to \textit{v2}, breaking changes will affect the client programs. Moreover, we ignore \textit{major upgrades} and \textit{developments} because these library upgrades can introduce breaking changes while \textit{non-major upgrades} need to ensure that there are no incompatible changes. We analyze the impact of breaking changes on client programs by calculating the usage related to breaking changes.

\subsubsection{Results}

According to Table \ref{tab:BCInUpgrade}, there are 17,009 \textit{non-major upgrades} and we collected 151,589 downstream client programs based on the dependency graph. We present the data of breaking changes used in client programs and library upgrades affected by different types of breaking changes, as shown in Table \ref{tab:distribution}.

There are 11,034 breaking changes used in client programs, in which the five most common breaking change types are \textit{element change} in \textit{named} type (33.5\%), \textit{add interface method} in \textit{interface} type (23.5\%), \textit{remove} in \textit{named} type (7.5\%), \textit{remove} in \textit{function} type (7.3\%), and \textit{param change} in \textit{function} type (7.0\%). Compared to the five most common breaking changes, \textit{element change} in \textit{named} type is added while \textit{value change} in \textit{basic (const)} type is missing. \textit{Named} types are data types customized by the developers of TPLs to achieve different functionality, therefore, breaking changes on them are more likely to be introduced by client programs. We ignore the missing of \textit{value change} in \textit{basic (const)} type because it still occurs frequently (5.3\%).

we also found that 96.9\% of breaking changes do not affect client programs. Among the high frequency of breaking changes, only a very low rate affects client programs, for example, only 0.6\% of \textit{remove} in \textit{function} type have an impact on client programs. For \textit{element change} in \textit{named} type, although its breaking changes occur at a rate of 2.8\%, one-third of them will have a significant impact.

We also explain the impact of breaking changes from the perspective of the client programs. There are 50,485 client programs affected by breaking changes, which means that one-third of the client programs should make more efforts due to violations of SemVer compliance. Moreover, the two most common breaking changes affecting client programs are \textit{add interface method} in \textit{interface} type (45.4\%) and \textit{value change} in \textit{basic (const)} type (37.2\%). \textit{add interface method} in \textit{interface} type requires the developers of client programs to write additional code to adapt to the new behaviour. And \textit{value change} in \textit{basic (const)} type indicates that the developers of client programs need to keep up with the times using newer parameters and configurations for program operation.

\begin{center}
\noindent\fbox{
\parbox{0.45\textwidth}{
\textit{Summary:} We observe that only 3.1\% of breaking changes are used by client programs, but they affect 33.3\% of downstream client programs due to numerous dependencies. This tells us that just a small number of breaking changes can have a serious impact on client programs in the Go ecosystem.
}
}
\end{center}

\section{Discussion}

\subsection{Implications for TPLs Developers}

Changes often occur in software development to add new features, fix bugs, and refactor code \cite{laerte2017historical}. When breaking changes exist, developers cannot perform \textit{minor upgrades} and \textit{patch upgrades} according to SemVer \cite{web:semver}. This is simple to do for TPLs maintained by s single or a small number of developers, while in a collaborative TPL, there is no guarantee that everyone will strictly adhere to SemVer compliance. Therefore, it is important for developers to use \textit{GoSVI} to detect breaking changes and help determine the version number.

Not all exported objects should be referenced by client programs. Developers can limit the scope of packages by defining \textit{internal} packages, which cannot be used by client programs. Also, if the interfaces do not want to be used by client programs, an unexported method should be added to the interface so that client programs cannot implement them.

Moreover, it makes sense to use \textit{GoSVI} to detect which exported objects are more popular with client programs, enabling developers to refactor their code more sensibly while affecting client programs as little as possible.

\subsection{Implications for Client Programs Developers}

As shown in RQ1 and RQ3, although 86.3\% of library upgrades adhere to SemVer compliance, 33.3\% of client programs are affected by breaking changes. Therefore, it is essential for developers of client programs to decide which version to upgrade. Compared to reading the documentation and source code, \textit{GoSVI} can automatically identify breaking changes of library upgrades and help meet requirements without causing too many interruptions.

Furthermore, breaking changes exist in most ecosystems \cite{alexandrel2021what}, such as Cargo, Npm, Packagist, and Rubygems, and Go is no exception. If developers do not want to be affected by breaking changes, the most conservative approach is to not update dependencies. Delaying updates to dependencies and checking feedback to decide whether to update is also a recommended approach.

\section{Threats To Validity}

\subsection{Internal Validity}

We collect the dataset from \textit{2018/09} to \textit{2023/04}, while SemVer was propused in \textit{2013/06}, meaning that the analysis results are missing between \textit{2013/06} to \textit{2018/09}. We ignore those data to ensure that the dataset is confident because it is not possible to extract dependency graphs and identify TPLs based on the module system.

We filter packages designed not to be imported based on the layout of Go project \cite{web:layout}, but this rule is not followed by all developers. Therefore, this threat may lead to a low rate of compliance with the Semver.

Although the Go apidiff tool is officially implemented and widely used, we were unable to verify its accuracy in detecting breaking changes because of the absence of any standard benchmark. In order to understand as much as possible about the effectiveness of \textit{GoSVI} in detecting breaking changes, we verified the precision on 100 randomly selected cases of breaking changes. The result \footnote{The data is accessible at https://drive.google.com/drive/folders/1Cf9KITH
z5p04xZJCkQQo5BZEP6h4Bov8} shows that \textit{GoSVI} can identify the majority of breaking changes with 91\% precision.

The process of resolving identifier nodes to extract the types may have errors. However, to relieve this threat, we implement this functionality based on a well-known TPL: \textit{packages} \cite{web:packages}.

\subsection{External Validity}

We collect the Go repositories hosted on GitHub, which attracts over 100 million software developers. There is an error in using the analysis results of repositories on GitHub as the analysis results of the whole ecosystem.

\section{Related Work}

\subsection{Study of Semantic Versioning Compliance}

SemVer is commonly accepted by package management systems to inform developers whether releases of software packages introduce possible breaking changes. Hence, many empirical words \cite{filiperoseiro2021anempirical,lina2022breaking,steven2017semantic} have studied compliance of SemVer in different package management systems, such as Java with Maven, Node.js with Npm, Rust with Cargo, PHP with Packagist and Ruby with Rubygems. Cogo et al. \cite{filiperoseiro2021anempirical} scoped the downgrade from client view with more than 600K reusable packages of JavaScript from Npm and the result showed the clients decide to downgrade mostly because the provider version introduces a failure and the downgrade is correlative with the upgrade caused by the provider. Ochoa et al. \cite{lina2022breaking} scoped SemVer on Java with Maven of almost 120K library upgrades and found that 20.1\% of non-major upgrades contained Breaking Changes and only 7.9\% of client programs are affected by breaking changes. Raemaekers et al. \cite{steven2017semantic,raemaekers2014semantic} scoped SemVer on Java with Maven of more than 100K libraries and found that nearly one-third of all releases introduce at least one breaking change, the tendency of complying with SemVer in non-major releases do not become better over time and developers of client programs do not follow deprecation guidelines suggested by SemVer as expected. In this paper, we focus on the scope of the Go package management system and found that 86.3\% of library upgrades comply with SemVer while 33.3\% of client programs may be affected by breaking changes.

\subsection{Detection of Breaking changes}

There are many tools that are designed to detect API breaking changes \cite{aline2018apidiff,mezzetti2018type,suhaib2020using,web:japicmp,zhang2022has}. Brito et al. \cite{aline2018apidiff} proposed APIDiff to detect syntactic changes on Java with type, method and field and recognized breaking changes and non-breaking changes based on different conditions. Mezzetti et al. \cite{mezzetti2018type} proposed techniques named type regression testing to automatically detect breaking updates on the public interface on Node.js based on dynamic analysis. Mujahid et al. \cite{suhaib2020using} proposed a technique to detect breakage-inducing versions of third-party dependencies. Moreover, Jezek et al. \cite{kamil2017api} have developed a compact benchmark data set of less than 200KB to evaluate the accuracy of the tools, such as japicmp \cite{web:japicmp}, JAPICC \cite{web:JAPICC}, and SigTest \cite{web:SigTest}. Zhang et al. \cite{zhang2022has} detected semantic breaking change based on static analysis and found that there were 2-4 times more semantic breaking changes than those with signature-based issues. In this paper, we focus on the signature-based breaking changes and propose \textit{GoSVI} tool to detect breaking changes between versions in the Go module system.

\subsection{Analysis of breaking change impact}

There are many works have been proposed to analyze the impact of breaking change on client programs \cite{laerte2017historical,steven2017semantic,lina2022breaking, robbes2012developers,Bavota2013Evolution}. Xavier et al. \cite{laerte2017historical} and Bavota et al. \cite{Bavota2013Evolution} analyzed whether the types of breaking changes are imported by client programs (using keyword \textit{import}). This approach overestimates the impact of breaking changes because there are cases where they are imported but not used. Robbes et al. \cite{robbes2012developers} manually calculated the ripple effects of breaking changes across an entire ecosystem. This approach is not appropriate for large-scale studies. Raemaekers et al. \cite{steven2017semantic} proposed a new technique to inject each breaking change individually and analyze its impact by counting the number of compilation errors. It is difficult to ensure that the injection process does not introduce other breaking changes to mislead the experimental results. Ochoa et al. \cite{lina2022breaking} used Rascal $M^3$ model \cite{Basten2015M3}, which stores the relationship between Java elements, to link breaking changes to the usages in client programs. This approach is limited by the $M^3$ model and cannot analyze breaking changes related to overridden methods. Prior works have focused on analyzing the impact of breaking changes in the Java language except for the Go language. In this paper, we resolved all identifiers in the client programs to extract their types and calculated the affected elements of client programs by comparing the types with breaking changes. We implement this functionality based on the Go underlying toolkit \textit{packages} \cite{web:packages} to make the results more accurate.

\subsection{API Evolution}

Many research works \cite{maxime2021asystematic, Rediana2019classification, robbes2012developers, Espinha2014web, sohan2015case, andré2014apievolutionminer, caroline2019what} have studied the API evolution from different perspectives. Lamothe et al. \cite{maxime2021asystematic} found that understanding, mitigating, and leveraging the impact of APIs and API evolution on software development are necessary to design and use APIs. Ko{\c{c}}i et al. \cite{Rediana2019classification} summarized that API evolution can be divided into changed API elements, changed performing actions from the producer of view, and changed API elements from the consumer of view. The reason to perform changes includes improving understandability and readability, improving error tolerance from various consumers, improving security by fixing bugs or meeting consumers' common requirements. Robbes et al. \cite{robbes2012developers} point out that the median time of developers reacting to API changes is two weeks, and the time of adaption takes nearly a month or higher. Hora et al. \cite{andré2014apievolutionminer} showed that client reaction time to API changes is longer than deprecation. Lima et al. \cite{caroline2019what} analyzed that popular APIs provide a large proportion of API elements and contain more code comments, which indicates the stability of popular APIs is better. In this paper, we aim to investigate API compatibility in terms of API breaking changes according to SemVer \cite{web:semver}.

\section{Conclusion}

In this paper, we conduct the first large-scale empirical study in the Go ecosystem to investigate the compliance of the SemVer in terms of breaking changes. We aim to analyze the frequency of breaking changes, the tendency of breaking changes over time, and the impact of breaking changes on client programs. To do so, we implement and use \textit{GoSVI} to detect breaking changes between two versions in the Go modules and analyze the impact of breaking changes on clients by resolving identifier nodes and comparing the types with breaking changes. Also, we collect a large-scale dataset with a dependency graph, including 124K TPLs and 532K client programs. We found the following results.

\begin{itemize}
    \item The number of library upgrades is rapidly increasing over time, and it is significant to explore whether each library upgrade adheres to Semver compliance.
    \item Library upgrades often introduce breaking changes and 28.6\% of \textit{non-major upgrades} violate SemVer compliance. Moreover, \textit{remove} in \textit{function} type is the most common breaking change condition at 36.2\%.
    \item More library upgrades do not introduce breaking changes over time. The tendency to comply with SemVer has significantly increased from 63.7\% in \textit{2018/09} to 92.2\% in \textit{2023/03}.
    \item There are only 3.1\% of breaking changes used in client programs, but one-third of client programs may be affected, meaning that few breaking changes will have a serious impact on client programs. 
\end{itemize}

According to our results, we give some suggestions to help developers of TPLs and client programs to make decisions about SemVer. Also, we recommend using \textit{GoSVI} to assist with automated breaking changes analysis and downstream client programs data insight.

As future work, we plan to detect semantic breaking changes in the Go ecosystem to understand how breaking changes potentially disrupt client programs. Also, we will draw comparisons with other package ecosystems such as Npm and Maven to provide more valuable insights into the relative significance of breaking changes across different platforms.

\section*{ACKNOWLEDGMENT}

We appreciate the insightful insights provided by anonymous reviewers to improve the quality of the paper. This work is supported by the National Science Foundation of China under grant No. 62072200 and No. 6217071437.

\bibliographystyle{IEEEtran}
\balance
\bibliography{references}

\end{document}